\documentclass[aps,prl,11pt]{revtex4}
\usepackage{amsfonts}
\usepackage{amssymb}
\usepackage{graphics}
\usepackage{graphicx}

\begin{document}

\begin{center}
\bigskip {\LARGE \textbf{Families of Bragg-grating solitons in a
cubic-quintic medium} } \\[0pt]
\vspace{12mm} {\large \textrm{Javid Atai$^{1}$ and Boris A. Malomed$^{2}$}} 
\\
\vspace{3mm} 
$^{(1)}${\it School of Electrical and Information Engineering, The
University of Sydney, Sydney, NSW 2006, Australia } \\
\vspace{1.5mm} 
$^{(2)}${\it Department of Interdisciplinary Studies,
Faculty of Engineering, Tel Aviv University, Tel Aviv 69978, Israel} \vspace{%
1.5mm} 
\end{center}

\vspace{18mm}

\section*{Abstract}

\vspace{0.5cm}

We investigate the existence and stability of solitons in an optical
waveguide equipped with a Bragg grating (BG) in which nonlinearity contains
both cubic and quintic terms. The model has straightforward realizations in
both temporal and spatial domains, the latter  being most realistic. Two
different families of zero-velocity solitons, which are separated by a
border at which solitons do not exist, are found in an exact analytical
form. One family may be regarded as a generalization of the usual BG
solitons supported by the cubic nonlinearity, while the other family,
dominated by the quintic nonlinearity, includes novel ``two-tier'' solitons
with a sharp (but nonsingular) peak. These soliton families also differ
  in the parities of their real and imaginary parts. A stability
region is identified within each family by means of direct numerical
simulations. The addition of the quintic term to the model makes the
solitons very robust: simulating evolution of a strongly deformed pulse, we
find that a larger part of its energy is \emph{retained} in the process of
its evolution into a soliton shape, only a small share of the energy being
lost into radiation, which is opposite to what occurs in the usual BG model
with cubic nonlinearity. \\
PACS: 42.81.Dp, 42.65.Tg, 42.81.Qb

\newpage

\section{\protect\bigskip Introduction}

It is commonly known that solitons in various physical media are supported
by balance between nonlinearity and dispersion. In particular, in the case
of the nonlinear Schr\"{o}dinger (NLS) equation, the cubic self-focusing
nonlinearity, which is most typical in dielectric optical media \cite
{Agrawal}, should be balanced by diffraction or anomalous chromatic
dispersion, in order to provide for the existence of spatial or temporal
solitons, respectively. If the relative sign of the diffraction/dispersion
and nonlinearity is opposite, solitons (bright ones) cannot exist.

The above arguments apply to intrinsic diffraction and dispersion in any
optical material. In some media, however, the intrinsic chromatic dispersion
is too weak to support solitons in the temporal or spatiotemporal domain;
then, much stronger artificial dispersion can be induced by means of a Bragg
grating (BG) written on the sample. Notable examples are temporal \cite
{DiTrapani} and spatiotemporal \cite{Wise} solitons that were created in
second-harmonic-generating crystals. In this case, the
size of available samples is a few cm, hence the corresponding soliton's
dispersion length must be very small, $\lesssim 1$ cm, which cannot be
provided for by the material's intrinsic dispersion, even if the soliton is
very narrow, with the temporal width $\,\sim 100$ fs. As  was
experimentally demonstrated in the above-mentioned works \cite{DiTrapani}
and \cite{Wise} (and proposed in the theoretical work \cite{Drummond}), the
necessary strong dispersion can be generated by BG. A similar situation
takes place in BG-carrying silica fibers, where very strong dispersion
induced by BG makes it possible to generate solitons (supported by the usual
Kerr nonlinearity) in a very short piece of  fiber \cite{Eggleton}.

The presence of BG gives rise to a dispersion relation between the frequency 
$\omega $ and propagation constant (wavenumber) $k$, $\omega ^{2}=\omega _{0}^{2}+k^{2}$, with the spectral gap $%
0<|\omega |<\omega _{0}$, inside which families of \emph{gap solitons} may
exist \cite{deSterke}. It can be readily seen that the expansion of the
dispersion relation near $k=0$ yields $\omega \mp \omega _{0}=\pm
\,k^{2}/\left( 2\omega _{0}\right) +...$, i.e., it contains two branches
with opposite signs of the dispersion. This simple argument implies that,
unlike media dominated by the intrinsic material dispersion, there should be
no essential difference between self-focusing (SF) and self-defocusing (SDF)
nonlinearity when the dispersion is induced by BG. Note that a
similar situation takes place for spatial solitons in a nonlinear planar
waveguide, if strong artificial diffraction is induced by BG in the form of
a system of parallel scores on a surface of the waveguide: in that case, the
same dispersion relation is valid, with $\omega $ and $k$ realized as the
longitudinal and transverse wavenumbers (see, e.g., Ref. \cite{Mak}).

The lack of principal difference between the SF and SDF nonlinearities
in BG systems suggests a possibility to introduce a model in which the
dispersion (or diffraction) is induced by BG, while the nonlinearity may
change its sign with the increase of the input power. Then, one may expect
that the model may give rise to two qualitatively different families of gap
solitons: low-frequency ones, in which the SF nonlinearity is balanced by
the dispersion branch with a sign corresponding to anomalous dispersion, and
high-power solitons, supported by the balance between SDF nonlinearity and
the normal branch of the dispersion. The simplest model of this type may be
based on the cubic-quintic (CQ) nonlinearity, that has recently attracted
considerable attention, chiefly in the multidimensional case, as the
combination of the SF cubic and SDF quintic terms prevents collapse and
makes it possible to anticipate the existence of stable solitons \cite{CQ}.
It has been shown that CQ nonlinearity correctly describes the dielectric
response of the \textit{PTS} crystal \cite{PTS}.

In this work, we introduce the BG-CQ model and study zero-velocity solitons
in it. The model is meaningful in both temporal and spatial domains, the
latter interpretation being most realistic. We find the solutions in an
exact analytical form, and their stability is tested by direct simulations.
We find, in accordance with the arguments given above, that the model gives
rise to two \emph{disjoint} families of stable solitons, one of which may be
regarded as a generalization of the ordinary BG solitons in the model with
the cubic nonlinearity \cite{exact}, while the other family consists of
essentially novel solitons, in which the quintic nonlinearity is dominating.

\section{The model and exact soliton solutions}

In the temporal domain, the model is based on a system of normalized coupled
equations for amplitudes $u(x,t)$ and $v(x,t)$ of the two counterpropagating
waves linearly coupled by the resonant reflection on BG: 
\begin{eqnarray}
iu_{t}+iu_{x}+\left[ (1/2)|u|^{2}+|v|^{2}\right] u-\nu \left[
(1/4)|u|^{4}+(3/2)|u|^{2}|v|^{2}+(3/4)|v|^{4}\right] u+v &=&0,  \label{u} \\
iv_{t}-iv_{x}+\left[ (1/2)|v|^{2}+|u|^{2}\right] u-\nu \left[
(1/4)|v|^{4}+(3/2)|v|^{2}|u|^{2}+(3/4)|u|^{4}\right] v+u &=&0,  \label{v}
\end{eqnarray}
where the BG-induced coupling coefficient is set to be $1$, and $\nu >0$ is
a real parameter to control the strength of the quintic nonlinearity.
Equations (\ref{u}) and (\ref{v}) assume the usual ratio $1:2$ of the
coefficients in front of the self-phase and cross-phase modulation terms in
the cubic part of the equation \cite{Agrawal}, and the ratio $1:6:3$ for the
quintic part was derived in Ref. \cite{Moscow}. The same equations apply to
the spatial domain, i.e., a planar waveguide with BG in the form of a system
of parallel scores, $x$ and $t$ being interpreted as the transverse and
longitudinal coordinates. In fact, this realization of the model is most
realistic, in view of the possibility to use the above-mentioned PTS crystal
that gives rise to the CQ nonlinearity.

In this work, we confine ourselves to the study of zero-velocity solitons,
which are sought for as 
\begin{equation}
u(x,t)=A(x)\exp (i\phi (x)-i\omega t),\,\,v(x,t)=B(x)\exp (i\psi (x)-i\omega
t),  \label{0V}
\end{equation}
where $\omega $ belongs to the above-mentioned gap, 
\begin{equation}
-1<\omega <+1,  \label{gap}
\end{equation}
and the real amplitudes $A,B$ and phases $\phi ,\psi $ satisfy equations
obtained by the substitution of (\ref{0V}) into Eqs. (\ref{u}) and (\ref{v}%
). After straightforward manipulations, two simple relations follow from
those equations, 
\begin{equation}
\frac{d}{dx}\left( A^{2}-B^{2}\right) =0,\,\frac{d}{dx}\left( \phi +\psi
\right) =0,  \label{constA}
\end{equation}
For the soliton solutions, the fields $A$ and $B$ must vanish at $x=\pm
\infty $, hence Eq. (\ref{constA}) yields $B(x)=A(x)$. A constant value of $%
\,\phi +\psi $ can be set equal to zero by means of an obvious phase shift,
hence we also have $\psi (x)=-\phi (x)$. Obviously, these relations between
the amplitudes and phases imply that the two fields are subject to the
relation $u^{\ast }=v$.

The remaining equations for the single amplitude $A$ and single phase $\phi $
take the following form: 
\begin{eqnarray}
\frac{da}{dx} &=&2a\sin (2\phi ),\,a\equiv A^{2},  \label{a} \\
\frac{d\phi }{dx} &=&\omega +\frac{3}{2}a-\frac{5}{2}\nu a^{2}+\cos (2\phi ),
\label{phi}
\end{eqnarray}
A quotient of Eqs. (\ref{phi}) and (\ref{a}) yields an equation for $\phi $
regarded as a function as $a$, 
\begin{equation}
\frac{dc}{da}+\frac{c}{a}=-\frac{3}{2}+\frac{5\nu }{2}a,\,c\equiv \omega
+\cos (2\phi ),  \label{c}
\end{equation}
which can be immediately solved to yield $c(a)=c_{0}/a-(3/4)a+(5\nu /6)a^{2}$%
, where $c_{0}$ is an arbitrary constant. Only the solution with $c_{0}=0$
is not singular, hence we finally have 
\begin{equation}
\cos (2\phi )=-\omega -(3/4)a+(5\nu /6)a^{2}.  \label{cos}
\end{equation}
The next step is to eliminate $\sin \phi $ from Eq. (\ref{a}) by means of
the expression (\ref{cos}), arriving at an equation 
\begin{equation}
\frac{da}{dx}=2a\sqrt{1-\left( \omega +\frac{3}{4}a-\frac{5\nu }{6}%
a^{2}\right) ^{2}}\,.  \label{dadx}
\end{equation}

An integral of Eq. (\ref{dadx}) can be written in an implicit form, 
\begin{equation}
x(a)=\int_{a}^{a_{0}}\frac{db}{2b\sqrt{1-\left( \omega +\frac{3}{4}b-\frac{5%
}{6}\nu b^{2}\right) ^{2}}}\,,  \label{int}
\end{equation}
which describes soliton solutions in the interval $0<x<\infty $ (for $x<0$,
the solution can be obtained trivially, as the function $a(x)$ is even).
Indeed, letting $a\rightarrow 0$ in Eq. (\ref{int}), one has $x\rightarrow
\infty $, in compliance with the boundary condition that $a(x)$ must vanish
at $|x|\rightarrow \infty $. On the other hand, it also follows from Eq. (%
\ref{int}) that $a(x)$ attains its maximum value $a_{0}$ (which is the
soliton's peak power) at $x=0$.

The integral in (\ref{int}) can be expressed in terms of incomplete elliptic
integrals, but this formal representation is useless. It is more essential
to find the soliton's peak power $a_{0}$ in terms of $\nu $ and $\omega $.
Because $da/dx=0$ at soliton's center, $a_{0}$ must be a root of the right
hand side of Eq. (\ref{dadx}), i.e., $\omega +(3/4)a-(5\nu /6)a^{2}=\pm 1$.
This pair of quadratic equations have four roots; after a simple
consideration, one can find that only two of them may be relevant, namely 
\begin{equation}
\left( a_{0}\right) _{1}=\frac{3}{5\nu }\left[ \frac{3}{4}-\sqrt{\frac{9}{16}%
-\frac{10\nu }{3}(1-\omega )}\right] ,  \label{1}
\end{equation}
\begin{equation}
\left( a_{0}\right) _{2}=\frac{3}{5\nu }\left[ \frac{3}{4}+\sqrt{\frac{9}{16}%
+\frac{10\nu }{3}(1+\omega )}\right] .  \label{2}
\end{equation}
Note that, in the limit $\nu \rightarrow 0$, the root (\ref{1}) yields the
peak power $(4/3)(1-\omega )$ of the Bragg-grating soliton in the usual
model with the cubic nonlinearity \cite{exact}, while the root (\ref{2})
goes to infinity in the same limit.

Further straightforward consideration of the structure of the exact solution
(\ref{int}) leads to a conclusion that, if the root (\ref{1}) exists as a
physical (real) one, i.e., in the region 
\begin{equation}
1-\omega <27/(160\nu ),  \label{<}
\end{equation}
this root determines the amplitude of the soliton, while the root (\ref{2})
is then irrelevant. Exactly at the border of the region (\ref{<}), $1-\omega
=27/\left( 160\nu \right) $, no soliton exists, and in the region 
\begin{equation}
1-\omega >27/(160\nu )  \label{>}
\end{equation}
the\ root (\ref{1}) does not exist. Nevertheless, Eq. (\ref{int}) yields a
soliton in the region (\ref{>}) too, but it takes a principally different
shape, with the amplitude given by the expression (\ref{2}).

Note that the solitons may exist only in the spectral gap (\ref{gap}), or $%
0<1-\omega <2$, therefore the region (\ref{>}) is actually present only if $%
\nu >27/320\approx 0.0\allowbreak 8\allowbreak 44$. Thus, in this case there
exist two different families of the exact soliton solutions separated by the
border $1-\omega =27/\left( 160\nu \right) $. Typical examples of solitons
of the two types are displayed in Fig. 1. In the region (\ref{<}), the
solitons may be regarded as obtained by a smooth deformation of the usual
gap solitons known in the model with the cubic nonlinearity \cite{exact},
while novel solitons existing in the region (\ref{>}), where the quintic
nonlinearity plays a dominant role, may have a characteristic ``two-tier''
structure, with a sharp (but nevertheless nonsingular) peak, as it is seen
in Fig. 1. Another qualitative difference between the solutions of the two
types pertains to their phase structure and parity: if the soliton's
amplitude takes the value (\ref{1}), it follows from Eq. (\ref{cos}) that $%
\cos (2\phi (x=0))=-1$, and the value (\ref{2}) of the amplitude corresponds
to $\cos (2\phi (x=0))=+1$. It is easy to understand (with regard to the
above normalization setting the constant value of $\phi (x)+\psi (x)$ equal
to zero) that a consequence of this is that, in the solutions of the first
type, \textrm{Re}$\mathrm{\,}u(x)$ and \textrm{Re\thinspace }$v(x)$ are odd,
\ and$\mathrm{\,}$\textrm{Im}$\,u(x)$\textrm{\ }and\textrm{\ Im}$\,v(x)$ are
even functions of $x$. In the solutions of the second type, the parities of
the real and imaginary parts of the solutions are opposite.

\section{Stability of the  solitons}

It is known that, even in the model with the cubic nonlinearity, the soliton
stability is a difficult issue. For the first time, a possibility of
instability of a part of the gap solitons was predicted in Ref. \cite{Tasgal}
on the basis of the variational approximation (VA). In that work, three
(quasi)modes of internal oscillations of a soliton were identified and  each
of which could become unstable, depending on the soliton's frequency $\omega $:
one mode generated weak nonoscillatory instability below a critical
frequency, $\omega <\omega _{\mathrm{cr}}^{(1)}$, and the others gave rise
to essentially stronger oscillatory instability in an interval $-1<\omega
<\omega _{\mathrm{cr}}^{(2)}$, with $\omega _{\mathrm{cr}}^{(2)}<\omega _{%
\mathrm{cr}}^{(1)}$. Later, the stability of the gap solitons in the cubic
BG model was studied rigorously by means of numerical methods applied to
equations for small perturbations linearized about the soliton \cite{Peli}.
As a result, it was found that a part of the solitons are unstable indeed.
Comparison with the predictions of VA shows that, while the above-mentioned
weak nonoscillatory instability appears to be an artifact of the
approximation (which can probably be explained by a general theory
considering possible false instabilities predicted by VA for solitons \cite
{Lakoba}), the stronger oscillatory instability sets in indeed if the
frequency $\omega $ is smaller than a critical value, which is very close 
to that predicted by VA.

By means of systematic simulations of Eqs. (\ref{u}) and (\ref{v}) we
 have tested the stability of both families of the solitons given by
 the exact implicit form (\ref{int}) against small initial 
perturbations. The resultant stability diagram in the
($\nu ,\omega $) plane is displayed in Fig. 2. The most important 
conclusion suggested by the diagram is that as $nu$ increases the 
stability region for the usual solitons (corresponding to solitons 
in region (\ref{<})) shrinks but the opposite occurs for the solitons 
of the novel type.

The simulations also demonstrate that far from the 
stability border the unstable solitons decay into 
radiation. However, in the vicinity of the 
stability border, after shedding some radiation, they 
rearrange themselves into stable solitons. Thus, with 
regard to the possibility of radiative losses, the 
stable solitons play the role of \emph{attractors} in 
the present conservative model, i.e., the stable solitons 
are really robust objects. A typical example is shown in 
Fig. 3, where an unstable soliton evolves into a stable one.

The  model presents essential differences against the  one
with pure cubic nonlinearity not only concerning the existence of solitons
and their stability against small perturbations, but also in dynamics of
strongly perturbed solitons. The simplest and most important example of a
strong perturbation is sudden uniform multiplication of the soliton's
profile by an amplification factor $\alpha $ essentially exceeding $1$,
which is a result of the action of an optical amplifier. In the usual models
with the cubic nonlinearity, it is well known that the amplified pulse
separates into a new soliton and considerable amount of radiation. However,
in the present model it is feasible that the quintic term, which has the
sign opposite to that in front of the cubic one, may help to shape the
amplified pulse into a new soliton, thus reducing the share of energy
emitted with radiation. Figure 4, which displays typical examples of the
relaxation of the strongly amplified (by the factor $\alpha =2$) pulses in
the cubic model proper and cubic-quintic one, demonstrates that this is
indeed the case. In fact, the effect of the quintic term (with $\nu =1/2$)
is very significant: in the cubic model, the final soliton retains only 
$11.6\%$ of the initial energy, while the energy-retention share in the
cubic-quintic model is $92.4\%$.

\section{Conclusion}

We have introduced a model whose linear part includes two counterpropagating
waves coupled by the resonant reflection on a Bragg grating, and the
nonlinear part includes cubic and quintic terms with opposite signs. The 
model gives rise to two different
families of solitons (provided that the coefficient in front of the quintic
term exceeds a minimum value, $\nu =27/320$), which are found in an exact
analytical form. One family may be regarded as a generalization of the usual
Bragg-grating solitons supported by the cubic nonlinearity, while the other
family, in which the quintic nonlinearity is dominant, includes ``two-tier''
solitons with a sharp (but nonsingular) peak. Also, the soliton families 
differ in the parities of their real and imaginary parts. In the
plane ($\omega ,\nu $), the two soliton families are separated by a curve on
which solitons do not exist. Stability regions have been identified within
each family by means of direct numerical simulations. It was also shown that
in the vicinity of the stability border the unstable solitons do not 
decay. Rather, after losing some energy in the form
of radiation, they evolve into stable ones. In the present model, almost all
the energy of a strongly perturbed soliton is retained in the process of its
evolution, a fairly small share being lost with radiation, which is opposite
to what occurs in the usual cubic model.

\newpage

\begin{center}
\textbf{Figure captions}
\end{center}

Fig. 1. Typical examples of solitons produced by the exact solution (\ref
{int}) in the regions (\ref{<}) and (\ref{>}) for $\nu =0.135$, shown,
respectively, by the dashed and solid curves. The two examples pertain to $%
\omega =-0.23$ and $\omega =-0.27$ (both are taken close to the border
between the two regions, $1-\omega =27/(160\nu )$). The inset shows a blowup
of the latter soliton near its center, to demonstrate that it has no
singularity at $x=0$.

Fig. 2. Stability diagram for both types of the solitons. The exact border
between the two families is the curve $1-\omega =27/\left( 160\nu \right) $,
see the text. To the left of the border (the region of the usual solitons),
both stable and unstable regions are explicitly marked. The novel solitons
for which the quintic nonlinearity is dominant occupy the region to the
right of the border, where the white area represents unstable solitons.

Fig. 3. An example of the evolution of an unstable soliton (with $\nu =0.1$%
,\thinspace $\omega =-0.1$) into a stable one.

Fig. 4. Evolution of a soliton pulse instantaneously multiplied by the
amplification factor $\alpha =2$: (a) the usual cubic model, $\nu =0$; (b)
the cubic-quintic model, $\nu =1/2$. In both cases, the original stable
stationary soliton pertains to $\omega =0.8$.\newline

\section*{}
\begin{figure}[th]
\centering{\ \includegraphics*[height=7cm,width=7cm]{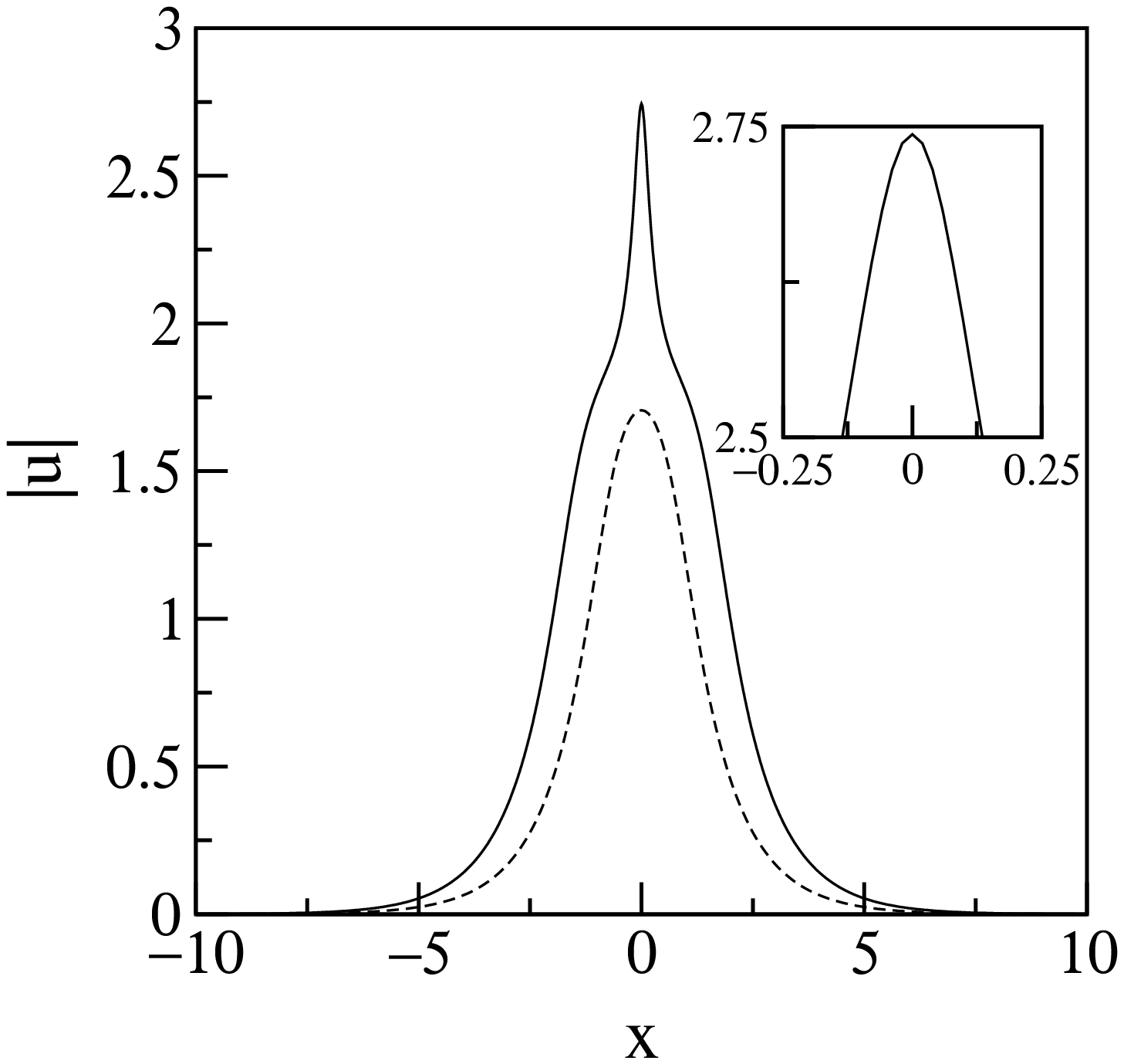}\\[2ex]
Fig.~1.~Javid Atai and Boris A.~Malomed}
\end{figure}
\pagebreak

\section*{}
\begin{figure}[th]
\centering{\ \includegraphics*[height=7cm,width=7cm]{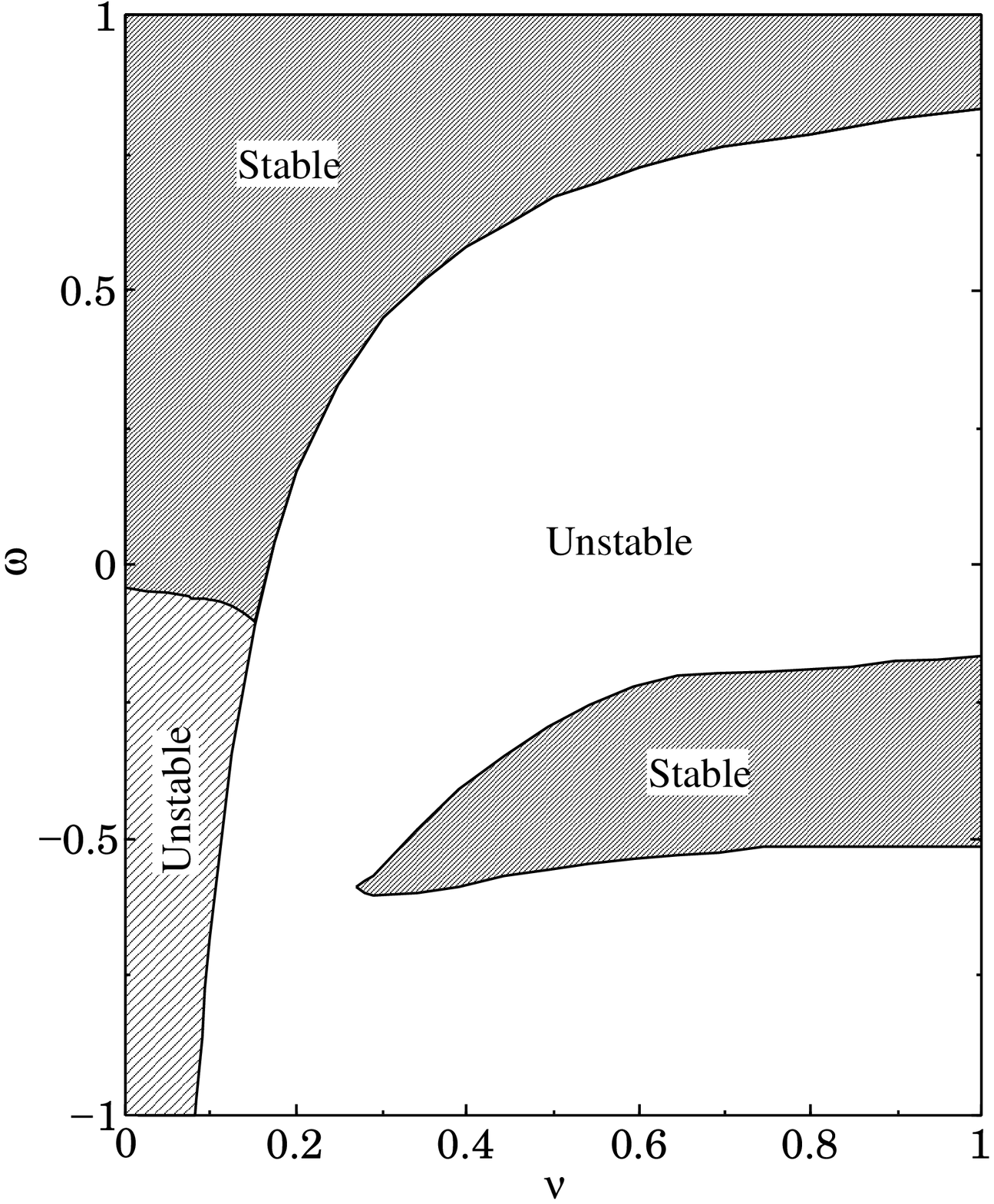}\\[2ex]
Fig.~2.~Javid Atai and Boris A.~Malomed}
\end{figure}
\pagebreak

\section*{}
\begin{figure}[th]
\centering{\ \includegraphics*[height=8cm,width=6cm]{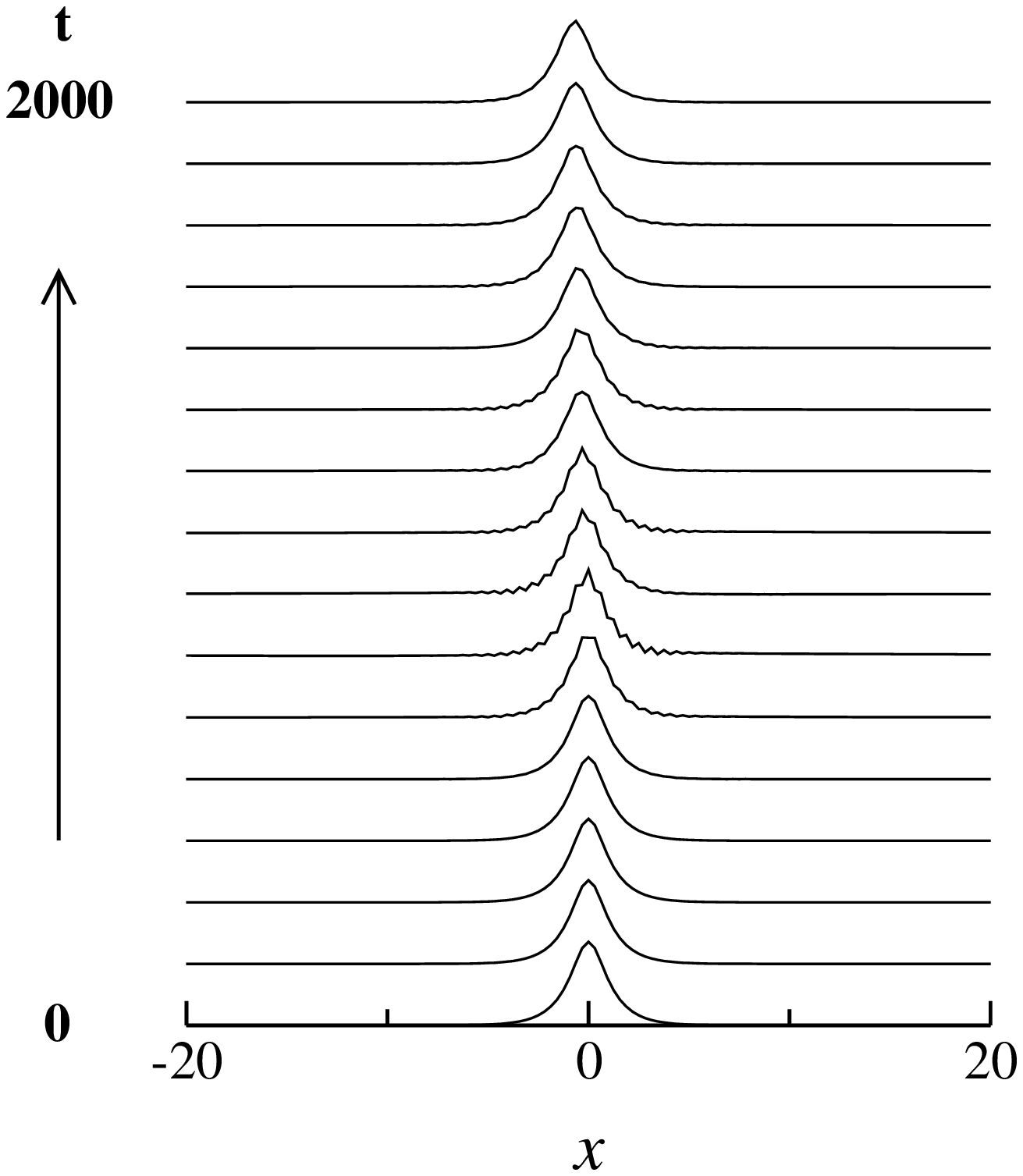}\\[2ex]
Fig.~3.~Javid Atai and Boris A.~Malomed}
\end{figure}
\pagebreak

\section*{}
\begin{figure}[th]
\centering{\ \includegraphics*[height=8cm,width=6cm]{fig4a.eps}\\[2ex]
Fig.~4.~Javid Atai and Boris A.~Malomed}
\end{figure}
\pagebreak

\section*{}
\begin{figure}[th]
\centering{\ \includegraphics*[height=8cm,width=6cm]{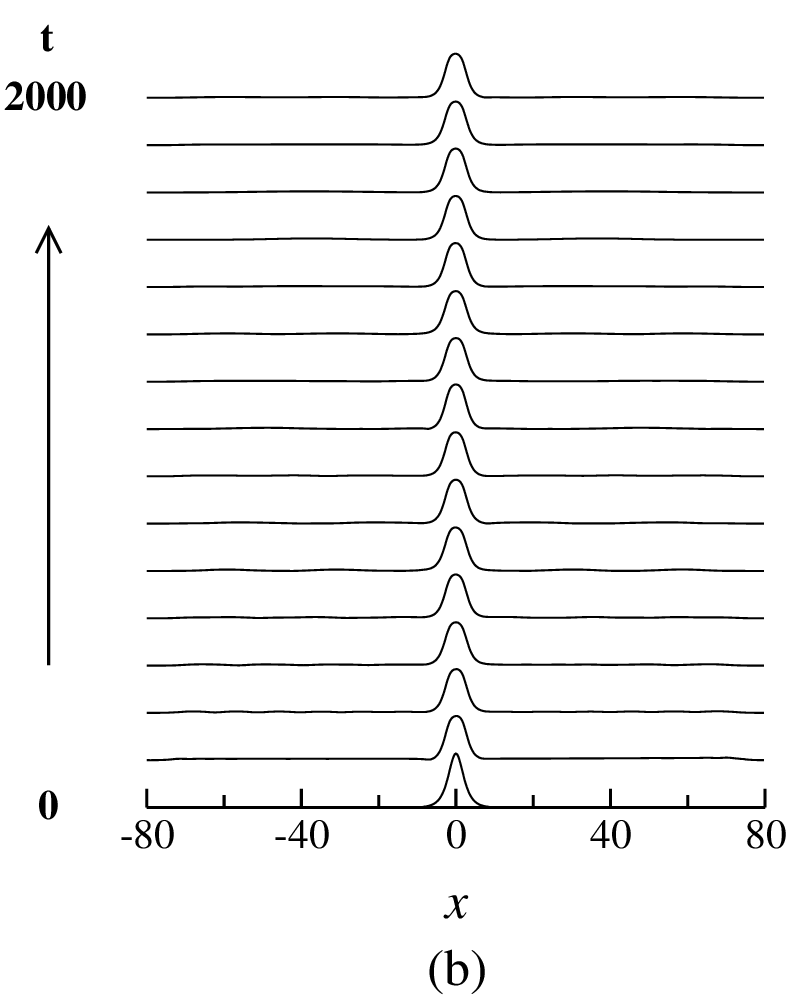}\\[2ex]
Fig.~4.~Javid Atai and Boris A.~Malomed}
\end{figure}

\end{document}